\documentclass[aps,prl,reprint,twocolumn,longbibliography]{revtex4-1}
\usepackage{amssymb}
\usepackage{amsmath}
\usepackage{bbold}
\usepackage{mathrsfs}
\usepackage{paralist}
\usepackage{amsmath}
\usepackage{graphicx}
\usepackage{color}
\usepackage{epstopdf}
\usepackage{subfigure}
\usepackage{color}
\usepackage{dsfont}
\usepackage{yfonts}
\usepackage{bbm}
\usepackage{mathrsfs}
\usepackage{bbold}
\usepackage{paralist}
\usepackage{xcolor}
\usepackage{bm}
\usepackage{pifont}
\usepackage{dcolumn}
\usepackage{latexsym}
\usepackage{xcolor}
\usepackage[%
  colorlinks=true,
  urlcolor=blue,
  linkcolor=blue,
  citecolor=blue
]{hyperref}
\usepackage{etoolbox}

\makeatletter
\let\cat@comma@active\@empty
\makeatother

\newcommand{\bra}[1]{\ensuremath{\langle #1 |}}
\newcommand{\ket}[1]{\ensuremath{| #1 \rangle}}

\newcommand{\inner}[2]{\ensuremath{\langle #1|#2 \rangle}}
\DeclareMathOperator{\sinc}{sinc}

\newcommand{\pr}{\prime}

\newcommand{\be}{\begin{equation}}
\newcommand{\ee}{\end{equation}}
\newcommand{\bgar}{\begin{eqnarray}}
\newcommand{\enar}{\end{eqnarray}}

\newcommand{\om}{\omega}

\begin{document}

\title{Generation Engineering of Heralded Narrowband Colour Entangled States.}

\author{A. Zavatta$^{1,2,3,^*}$, M. Artoni$^{3,4}$, G. La Rocca$^5$.}
\affiliation{$^1$ Istituto Nazionale di Ottica (CNR-INO) Largo E. Fermi, 6 - 50125 Firenze, Italy.}
\affiliation{$^2$ Department of Physics and Astronomy, Universit\`a di Firenze,  Italy.}
\affiliation{$^3$ European Laboratory for Nonlinear Spectroscopy, Firenze, Italy.}
\affiliation{$^4$ Department  of Engineering and Information Technology, Brescia University, Brescia, Italy.}
\affiliation{$^5$ Scuola Normale Superiore and CNISM, Pisa, Italy.}

\date{\today}

\begin{abstract}
Efficient  heralded generation of entanglement together with its manipulation is of great importance for quantum communications. In addition, states  generated with bandwidths naturally compatible with atomic transitions allow a more efficient mapping of light into matter which is an essential requirement for long distance quantum communications. Here we propose a scheme where the indistinguishability between  two spontaneous four-wave mixing processes is engineered to herald generation of single-photon frequency-bin entangled states, {\it i.e.} single-photons shared by two distinct frequency modes. We show that entanglement can be optimised together with the generation probability, while maintaining  absorption negligible. Besides,
the scheme illustrated for cold rubidium atoms is versatile and can be implemented in several other physical systems.
\end{abstract}

\pacs{Valid PACS appear here}

\maketitle

Quantum states of light are  crucial  for  quantum communications;
however, the propagation of quantum states through optical channels is dramatically affected by losses, and worldwide quantum networks remain a hard task to accomplish. In spite of this, the seminal paper of Duan {\it et al.}~\cite{Duan2001} proposed long distance quantum communications  through  the  entaglement of distant atomic ensembles. Since then, a growing attention has been  devoted towards the development of narrowband sources of quantum states of light to efficiently map light into atoms. Recently, several generation schemes based on spontaneous four-wave mixing   and Raman processes in atomic ensembles have  been  proposed and demonstrated such as, \textit{e.g.}, entangled photon pairs generated in trapped cold atoms~\cite{Kolchin2006,Srivathsan2013,Liao2014,Han2015} and in hot atoms vapor cells~\cite{Shu2016}. In such schemes, the quantum states  can be generated with bandwidths naturally compatible with the atomic transitions, allowing a more efficient mapping of light into matter. 
Unfortunately, such processes are probabilistic and the generation happens at random time with very low probability. In order to circumvent this limitation, so called heralded schemes, where the output states are freely available for further processing, without post-selection, are required to permit both their characterization with homodyne tomography~\cite{MacRae2012} and their use into a more complex scenario.

\begin{figure*}[th]

\includegraphics[width=13.5cm]{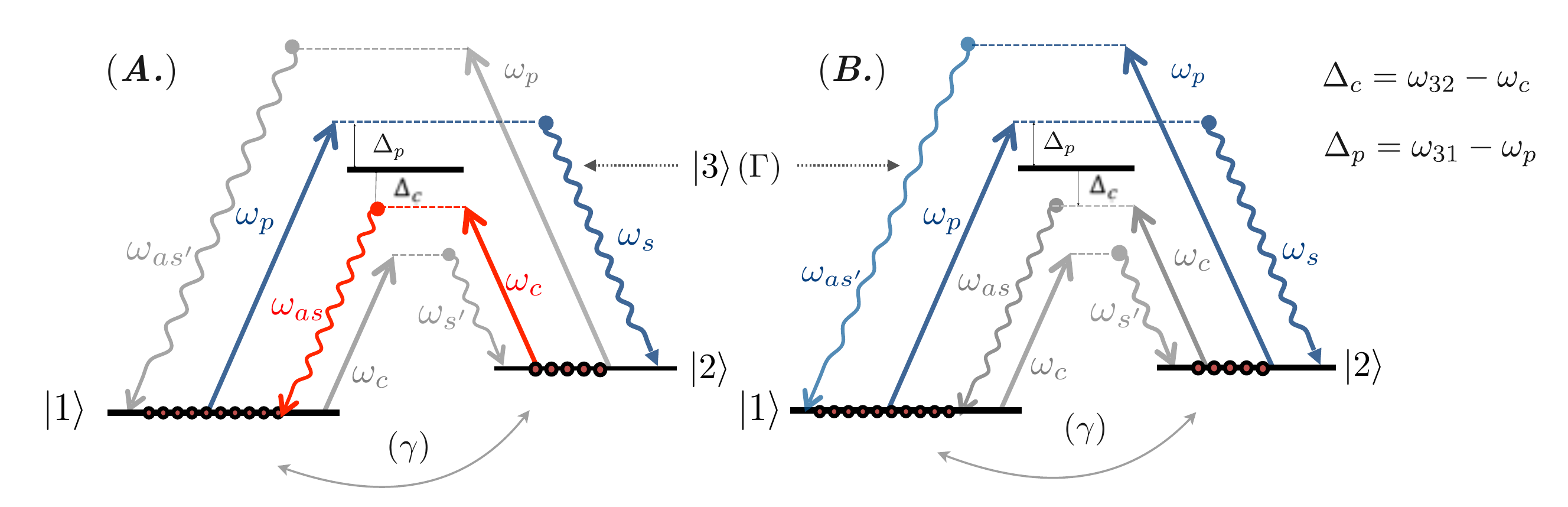}
\caption{
\label{fig:levels} 
\textbf{Single-photon color entanglement.}
[\textbf{\textit{A.}}]
A pair of  weak coupling ($\omega_c$) and pump  ($\omega_p$) beams spontaneously scatter through the third-order nonlinearity of a three-level atom to generate an \textit{anti-Stokes} photon $\omega_{as}$ (\textit{red}) and  a $Stokes$  photon $\omega_s$ (\textit{blue}).
\textbf{[\textit{B.}]} 
The different pair $\{ \omega_{as^\pr}, \omega_{s}\}$ of photons can be generated through spontaneous  scattering off the pump ($\om_p$) alone.
The spontaneous four-wave mixing processes \textit{A.} and \textit{B.} result in a small probability of generating single anti-Stokes photons at frequencies
 $\omega_{as}$ and $\omega_{as^\prime}$.
For certain atomic parameters ($\gamma, \Gamma, N/V$) and (driving) pump and coupling detunings ($\Delta_{p,c}$) and Rabi frequencies ($\Omega_{p,c}$)~\cite{sm}, the emission of one single-photon through channels \textit{A.} and \textit{B.} can be made to be \textit{undistinguishable} subject to a state-projection measurement on the Stokes ($\om_s$) photon  (Fig.~\ref{fig:exp-scheme}).
The sample has a density $N/V \simeq 5 \times 10^{12}$cm$^{-3}$, length $L=100 \mu$m, while atomic levels correspond to the transition 5$^{2}$S$_{1/2}$ $\to$ 5$^{2}$P$_{1/2}$ ($^{85}$Rb $D_1$ line) with $\lambda_{31}$=795 nm, $\omega_{21}=2 \pi \times 3$~GHz and decay rates 
 $\Gamma=2 \pi \times 5.75$~MHz  and $\gamma=2 \pi \times 10$ kHz.
}
\end{figure*}

In this context, it is interesting to exploit  the multilevel structure of a resonant media such as atoms or atom-like systems~\cite{Arimondo:2016aa}, in order to engineer the generation of quantum states. More specifically, in this Letter, we propose a scheme 
for the heralded generation of single-photon color entangled states \textit{i.e.},  states where a single photon excitation is simultaneously shared by two distinct frequency modes of light, exploiting indistinguishably between four-wave mixing processes.
This state may be envisaged as the frequency counterpart of the familiar single-photon path entanglement~\cite{Morin2013, Monteiro2015}, which can also be extended to  temporal domain~\cite{Zavatta2006} and used as a important resource in quantum protocols to convey and process quantum information~\cite{Lombardi2002, Salart2010, Chou2005, Choi2008}. 
The propagation of color-entangled  states through a specific atomic interface has also been studied~\cite{Zavatta2014}, together with their mapping into a quantum memory based on electromagnetically induced transparency technique in a double-$\Lambda$ media~\cite{viscor2012}. Therefore, the scheme proposed here represent  an additional building block to realize a quantum network, in which colour-entanglement  is generated, manipulated and stored.

\begin{figure}[t!]
\includegraphics[width=6cm]{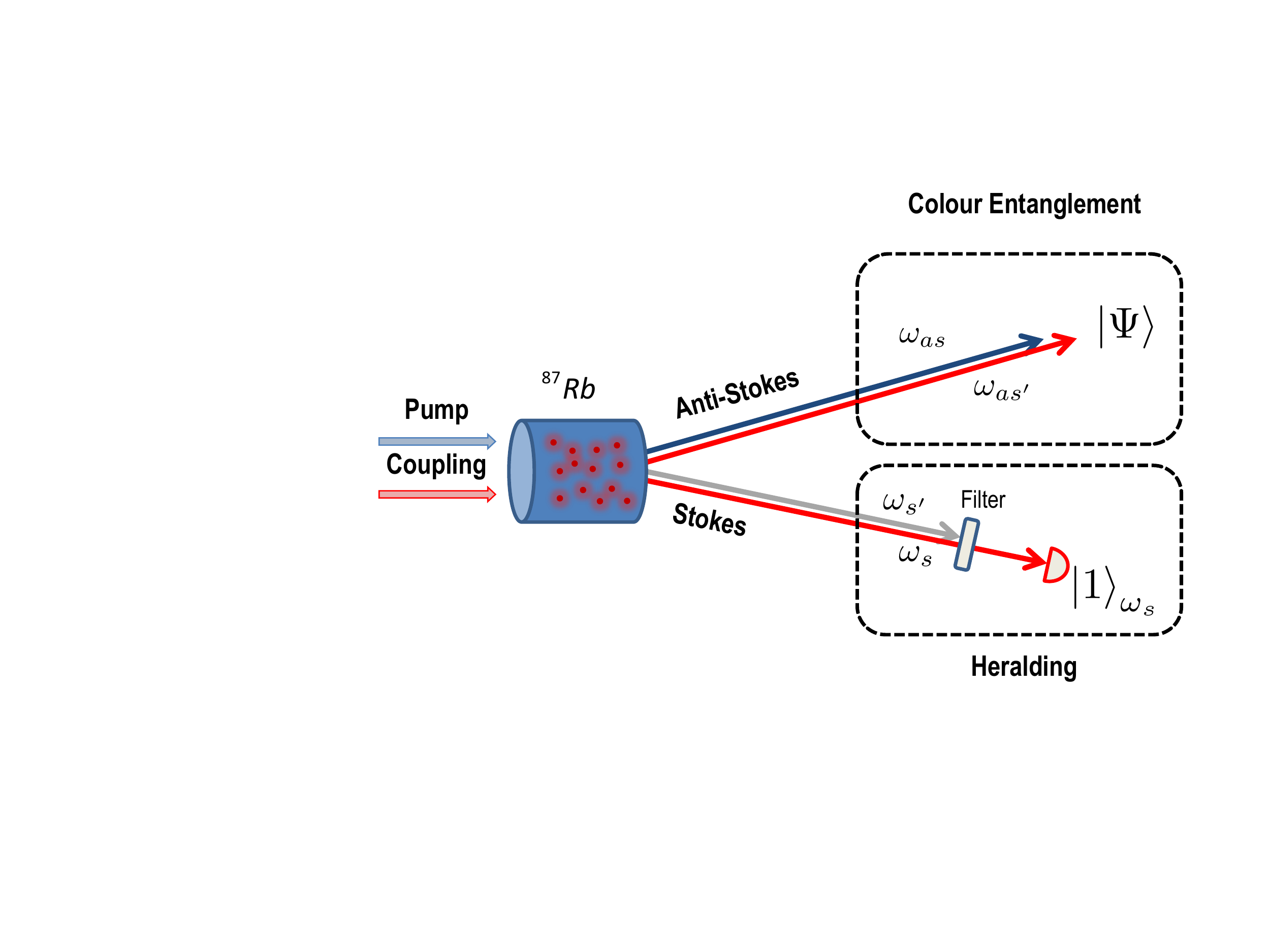} 
\caption{
\label{fig:exp-scheme} 
\textbf{\textit{Stokes state projection.}}
Under specific pump and coupling drivings of a sample of cold $^{85}$Rb atoms pairs of pump and coupling photons scatter via spontaneous four-wave mixing into pairs of Stokes and anti-Stokes photons 
$\{ \omega_s, \omega_{as} \}, \{ \omega_s, \omega_{as^\prime} \}$  $\{ \omega_{s^\prime}, \omega_{as} \}$ and $\{ \omega_{s^\prime}, \omega_{as^\prime} \}$. 
A "single" photon can be shared by two different modes $\omega_{as}$ and $\omega_{as^\prime}$ \textit{if} a  Stokes photon $\omega_{s}$ is detected 
(\textit{lower-box}).
Such a conditional detection ``heralds''  the generation of the entangled state $\ket{\Psi}$ (\textit{upper-box}).
\label{fig:scheme}
}
\end{figure}

Here we describe  a scheme where ``two'' specific four-wave mixing processes are exploited to engineer heralded  single-photon frequency-bin entangled states.
The scheme follows from a four-photons spontaneous four-wave mixing process in a three-level third-order nonlinear medium where, in the presence a pair of \textit{weak} coupling ($\omega_c$)  and pump ($\omega_p$) co-propagating beams, Stokes and anti-Stokes photon pairs emerge (Fig.\ref{fig:levels}).
The two processes, one (\textit{A}) leading to the emission  of a \textit{Stokes} ($\omega_s$) and an \textit{anti-Stokes} ($\omega_{as}$) photon pair and the other (\textit{B}) leading to the emission of the same \textit{Stokes} ($\omega_s$) yet a different \textit{anti-Stokes} ($\omega_{as^\prime}$) photon are sketched in Fig.~\ref{fig:levels}(\textit{colour lines}).
The main feature of the scheme is that  under suitable pump and coupling drivings 
(\textit{i}) the probability that the two processes \textit{A} and \textit{B} occur simultaneously can be made to be negligible while  (\textit{ii}) the generation of anti-Stokes photons  either through process \textit{A} or through process \textit{B} can be made to occur with the same probability.
Hence  the detection of a single-photon at frequency $\omega_s$  will entail, via a state-projection measurement (Fig.~\ref{fig:exp-scheme}),
the emission of an anti-Stokes photon though its detection does not reveal, 
even in principle, which photon $\omega_{as}$ or $\omega_{as^\prime}$ has been emitted. Such an indistinguishability results in the state \textit{superposition},
\begin{equation}
\ket{\Psi}
=
\alpha \ \ket{1}_{\omega_{as}}\ket{0}_{\omega_{as^\prime}}+\beta \ \ket{0}_{\omega_{as}} \ket{1}_{\omega_{as^\prime}}.
\label{state}
\end{equation}
It turns out that $\alpha$ and $\beta$ essentially depend on  the medium linear and nonlinear properties hence directly tunable through the pump and coupling driving fields.

Two more processes will clearly contribute to the heralded generation of single-photon state (\ref{state}), namely those involving the spontaneous generation of the different Stokes-anti-Stokes pairs $\{\omega_{s^\prime}, \omega_{as^\prime}\}$ (\textit{C}) and $\{ \omega_{s^\prime}, \omega_{as} \}$ (\textit{D})
(\textit{gray lines} in Fig.\ref{fig:levels}).
However the projection procedure above will prevent the processes \{\textit{C},\textit{D}\} from concurring to the generation of the state $\ket{\Psi}$~\cite{noAB}.

\textbf{\textit{Entanglement generation.}}
Assuming classical fields in the form of plane-waves
$E_{i}^{+} = E_{i} e^{i \textbf{\textit{k}}_{i} \cdot \textit{\textbf{r}}- i \omega_{i} t}$ with $i=\{c,p\}$, respectively for the control and pump beams, and standard field operator
\begin{equation}
\hat E_{j}^{+} = \frac{1}{\sqrt{\pi}} \int d\omega \sqrt{\frac{2\hbar \omega}{c \epsilon_0 A}}e^{i [\textit{\textbf{k}}_j(\omega) \cdot \textit{\textbf{r}}-\omega t]} \hat a_j(\omega)
\end{equation}
with $j=\{a,as\}$, for the Stokes and anti-Stokes photons propagating with a (complex) wave-vector $\textit{\textbf{k}}_j(\omega)$~\cite{loudon, Artoni:1999aa,  Artoni:1998aa},
the effective Hamiltonian describing the photon-atom interaction  can be written as (\textit{interaction picture}),
\begin{widetext}
\begin{eqnarray}
\label{ham}
\hat H_I
=
   \frac{\epsilon_0 A}{4} \int_{-L/2}^{L/2} dz \big[ \chi^{(3)}_A E_p^{+}  E_c^{+} \hat E_{as}^{-} \hat E_{s}^{-}
+ \chi^{(3)}_B E_p^{+}  E_p^{+} \hat E_{as'}^{-} \hat E_{s}^{-} 
+ \chi^{(3)}_C E_c^{+}  E_p^{+} \hat E_{as'}^{-} \hat E_{s'}^{-} 
+  \chi^{(3)}_D E_c^{+}  E_c^{+} \hat E_{as}^{-} \hat E_{s'}^{-} \big]
+ \textit{h.c.},
\end{eqnarray}
\end{widetext}
with the fields (negative) frequency parts $E_{i}^{-} $ and $\hat{E}_{j}^{-}$ computed as usual~\cite{loudon} and with space-time dependencies purposely omitted here.
The third-order optical nonlinear susceptibility $\chi^{(3)}_{l}$ for $l=\{A,B,C,D\}$ in (\ref{ham}) corresponds to the four spontaneous nonlinear mixing processes triggered by  coupling and pump.
Two of these processes are highlighted in color in Fig.~\ref{fig:levels}, while a shade of grey has been used there for the other two, with the full  susceptibilities expressions given in~\cite{sm}.
For the sake of clarity we restrict to a nearly \textit{1D} scattering geometry
 with Stokes and Anti-Stokes photons emitted in the $z$-direction and with a  (transverse) mode-profile cross-section $A$ across the interaction region $L$.
We further assume that the Stokes and the anti-Stokes are initially in the vacuum $\ket{0}_s$ $\ket{0}_{s'}\ket{0}_{as}$ $\ket{0}_{as'} \to \ket{0}$,
so that in the weak spontaneous scattering limit~\cite{Du2008} one has for the output state,
\begin{widetext}
\begin{eqnarray}
\ket{\Psi}_{out} &\simeq& \Big(\mathbf{1}  -  \frac{i}{\hbar} \int_{-\infty}^{\infty} dt\ \hat H_{I}\Big)\ket{0} \nonumber \\
&=&\ket{0} +   \int d \omega  \Big[
f_A(\omega_c+\omega_p-\omega,\omega)\hat a^\dag_{as}(\omega_c+\omega_p-\omega) \hat a^\dag_{s}(\omega) +
f_B(2\omega_p-\omega,\omega)\hat a^\dag_{as'}(2\omega_p-\omega) \hat a^\dag_{s}(\omega)   \nonumber \\
&+&    
f_C(\omega_p+\omega_c-\omega,\omega)\hat a^\dag_{as'}(\omega_p+\omega_c-\omega) \hat a^\dag_{s'}(\omega)+
f_D(2\omega_c-\omega,\omega)\hat a^\dag_{as}(2\omega_c-\omega) \hat a^\dag_{s'}(\omega)   \Big] \ket{0}_s\ket{0}_{s'}\ket{0}_{as}\ket{0}_{as'}.
\label{eq:emittedstate}
\end{eqnarray}
\end{widetext}
Although states with more than one photon per mode are possible due to higher order processes, they are nevertheless unlikely owing to small nonlinearities of the weak spontaneous scattering~\cite{Du2008} we examine here, hence the leading term (\ref{eq:emittedstate}) is adequate.
These four ``two-photon'' states are generated, in  general, with  \textit{different} amplitude probabilities $f_l$ for each of the four terms $l=\{A,B,C,D\}$ (Fig.~\ref{fig:levels}\textit{A}).
Specifically, generation of the  pair $\{\omega_s, \omega_{as}\}$, \textit{e.g.}, occurs with the probability amplitude,
\begin{widetext}
\begin{equation}
f_A(\omega,\omega') = - i \frac{\sqrt{\omega \omega'}}{4 \pi  c} \chi^{(3)}_A(\omega,\omega') E_p E_c \sinc\left(\frac{\Delta k_A L}{2}\right) L,
\label{eq:fa}
 \end{equation}
\end{widetext}
and depends on the susceptibility $\chi^{(3)}_{A}(\omega, \omega^{\prime})$ through  the electric field amplitudes ($E_p$,$E_c$) and on the  $z$-projection $\Delta k_A =(\textbf{\textit{k}}_{as} + \textbf{\textit{k}}_{s}- \textbf{\textit{k}}_p -\textbf{\textit{k}}_c)\cdot \hat{\textbf{\textit{z}}}$ of the wavevector mismatch (momentum-conservation) for co-propagating pump and coupling beams. We restrict here to collecting pairs of Stokes and anti-Stokes photons that are emitted with a very small angle with respect to the $z$-direction.
The  probabilities for the processes ${B,C,D}$ are instead obtained exchanging $c\rightarrow p$ and $as\rightarrow as'$ [to obtain $f_B(\omega,\omega')$],
$c\leftrightarrow p$, $s\rightarrow s'$ and $as\rightarrow as'$ [$f_C(\omega,\omega')$],
and $p\rightarrow c$ and $s\rightarrow s'$ [$f_D(\omega,\omega')$], while the same applies when computing the other mismatches $\Delta k_{B,C,D}$.
It may now be seen  from (\ref{eq:emittedstate}) that the detection of a Stokes photon at frequency $\omega_{s}$  (Fig.~\ref{fig:exp-scheme}) projects $\ket{\Psi}_{out}$ into a superposition of anti-Stokes single-photon with frequency $\omega_{as}$ and $\omega_{as'}$,
\begin{eqnarray}
\ket{\Psi}_{p}&=&\!\!\Big[  {_s}\bra{0}\hat a_s(\omega_s)\Big]\ket{\Psi}_{out}=\nonumber \\
&&\!\!\!\!\!\!\!\! f_A(\omega_{as},\omega_{s}) \ket{1}_{as}\ket{0}_{as^\prime}
+
f_B(\omega_{as'},\omega_{s})\ket{0}_{as} \ket{1}_{as^\prime}
\label{final}
\end{eqnarray}
in which corresponds to  $\ket{\Psi}$ (\ref{state}) with (complex) coefficients 
$\alpha \to f_A(\omega_{as},\omega_{s})/\surd {\cal{N}}$ and 
$\beta \to f_B(\omega_{as'},\omega_{s})/\surd {\cal{N}}$, being  ${{\cal{N}}}=|f_A |^2+|f_B|^2$ (state normalization). We used the notation  $\ket{1}_{i}=\hat a^\dag(\omega_i)\ket{0}$ to indicate a single-photon state at the frequency $\omega_i$ which defines the frequency mode $i=\{as,as^\prime,s,s^\prime\}$. 
The degree of entanglement in $\ket{\Psi}_p$,
which   depends through $ f_A$ and $f_B$ on the medium optical response (linear and nonlinear) and the wavevector mismatch  can then be all-optically controlled 
 through  pump and coupling  (Fig.~\ref{fig:levels}).
 It turns out infact that we can span continuously from a \textit{maximally} entangled state ($ \alpha = \beta = 1/\sqrt{2}$) to a \textit{pure} photon state~\cite{single-photon}
 ($ \alpha= 0 $ or $ \beta = 0$) directly through the Rabi frequencies $\Omega_{p,c}$ and  detunings $\Delta_{p,c}$~\cite{rel-phase}.
 We adopt here the negativity of the partial transpose ($NPT$) concept to quantify the degree of   entanglement. 
This has been introduced in~\cite{Lee2000} and it is based on the Peres-Horodecki \cite{PeresAsher1996,Horodecki1996} separability criterion thereby 
 $NPT=1$ corresponds to maximally entangled states and $NPT=0$ to separable ones. 
 For the specific state $\ket{\Psi}_p$ in (\ref{final}) we have $NPT = 2 |\alpha||\beta|$.

\begin{figure}[t!]
\includegraphics[width=5 cm]{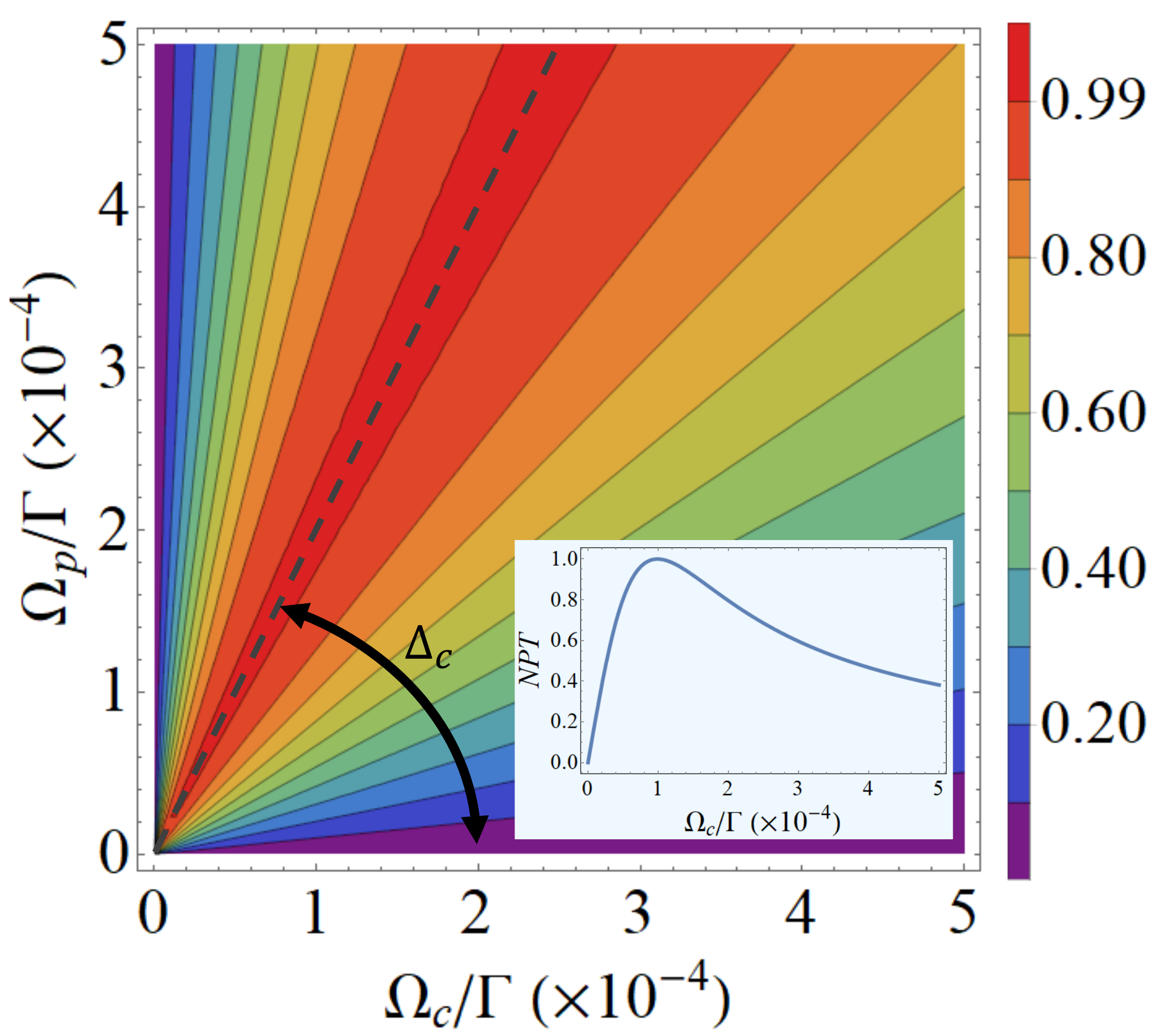}  
\caption{
{\it \textbf{Raman-like regime.}}
Negativity of the partial transpose ($NPT$) for the state $\ket{\Psi}$ in (\ref{state}) with
$\Delta_p = -1$ GHz, $\Delta_c = 10$ GHz vs. varying pump and coupling intensities ($\propto \Omega_{p,c}$).
For fixed driving intensities, maximum NPT values (\textit{dashed line}) may be attained by tuning $\Delta_c$.
\textit{\textbf{Inset.}} 
Plot section at $\Omega_p = 2 \times 10^{-4} \Gamma$:
maximum entanglement is tuned through the coupling intensity ($\propto \Omega_c$). 
\label{fig:raman-eit}
}
\end{figure}

\textbf{\textit{Entanglement engineering.}}
We will focus in the following on three representative situations to illustrate how entanglement can be efficiently engineered.
We choose  a  realistic set-up made of cold $^{85}$Rb atoms~\cite{Foot:2005aa}, though the generation mechanism is general enough to be suited to non-atomic architectures as well~
\cite{Venkataraman:2011ys, Bajcsy:2011yu,Doherty:2013uq, Rost:2010aa,He:2006aa,PhysRevB.77.113106}.
We start considering the case of weak driving fields and both far detuned from resonance,\textit{ i.e.}, 
$\Omega_{p,c} \ll \sqrt{\gamma \Gamma}$ and  $|\Delta_{p,c}| \gg \Gamma$.
This is reported in ~(Fig.~\ref{fig:raman-eit}) where a nearly balanced spontaneous Raman scattering regime with $NPT = 1$ in a parameters region for which $\Omega_c/\Delta_c \sim \Omega_{p}/\omega_{21}$.
Maximum  entanglement can then be directly tuned through the off-resonant coupling detuning ($\Delta_c$) (\textit{dash-line}), else through its intensity  ($\Omega_c$) (Fig.~\ref{fig:raman-eit} inset), and can always be achieved at vanishing levels of absorption ($\sim 10^{-4}$).
The generation probability of the entangled state is the probability to find a single-photon state in modes $\omega_{as}$ and $\omega_{as'}$ and this is ${\cal{P}}= |\inner{1_{as}}{\Psi}_p|^2+|\inner{1_{as'}}{\Psi}_p|^2=|f_A |^2+|f_B|^2$. 
In the spontaneous Raman regime values of $NPT$ close to unity are attained with  probabilities ${\cal{P}}\sim 10^{-13}$  when background level populations  are completely unbalanced ($\rho_{11}^0  \gg \rho_{22}^0$).

Anti-Stokes generation efficiencies ($\propto {{\cal{P}}}$) will improve  \textit{if} one
or both driving fields are brought close to resonance though absorption at each anti-Stokes mode will become important~\cite{Artoni:1999aa}.
This turns out to be an important generation regime which we will discuss below.
A proper assessment of the negativity of the partial transpose function in the presence of absorption is in order, which is done now by adopting a standard beam splitter model~\cite{kiss1995}. 
According to this model,  the dissipative process is formally described  by means of a beam-splitter that mixes the ideal state with the vacuum, then the partial trace over the lossy channel returns the state subjected to losses. Thus, the lossy state appears to be noisy and attenuated by the transmission coefficient of the beam splitter.

We apply this losses model  to each anti-Stokes $i=\{as,as'\}$ mode with transmission $T_i (\omega) $, which is obtained from  the imaginary part of the first order susceptibility $\chi_i^{(1)}(\omega)$~\cite{sm}, $T_i(\omega)=\exp\{-\mbox{Im}[\chi_i^{(1)}(\omega)]\omega L/c\}$. The output  state affected by absorption  is then used to calculate the $NPT$. Hereafter, a $NPT$ close to unity will then ensure the generation of a maximally entangled state with negligible absorption.

\begin{figure}[t!]
\includegraphics[width=5cm]{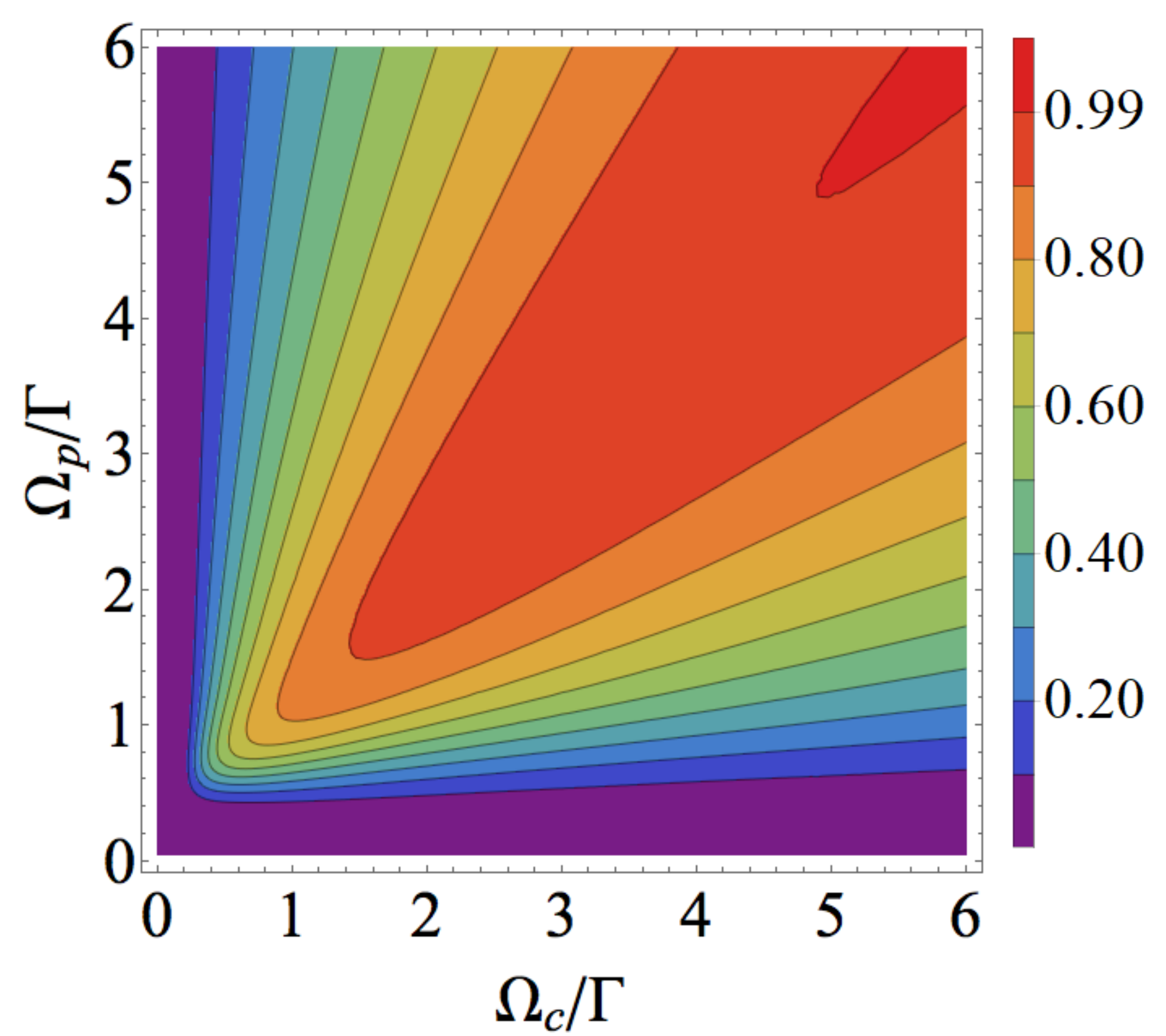} 
\caption{
{\it \textbf{Near-resonance regime.}}  
Same as in Fig.~\ref{fig:raman-eit} with $\Delta_p=-\Gamma$, $\Delta_c=\Gamma$.
The generation probability  is almost constant ${\cal{P}}\sim 10^{-5}$ in the region in which $NPT>0.9$, while increasing as $NPT$ decreases. 
The anti-Stokes $\omega_{as}$ losses are constant with $\Omega_p$ yet varying with $\Omega_c$ from total absorption to $\lesssim $ 0.2\% ($\Omega_c=6 \ \Gamma$).
Anti-Stokes $\omega_{as'}$ losses,
constant with  $\Omega_c$, 
range from a $\sim$ 3\% max ($\Omega_p\simeq \Gamma$) to almost no loss at all ($\Omega_p = 6 \ \Gamma $).
\label{fig:eit1}
}
\end{figure}

We report in Fig.~\ref{fig:eit1} the $NPT$ behavior for small detunings $|\Delta_{p,c}|\sim\Gamma$ and driving fields such that $\sqrt{\gamma\Gamma} \lesssim \Omega_{p,c} \sim \Gamma$, showing   large variations of the  degree of entanglement along with losses and occurring with generation probabilities ${{\cal{P}}}$ orders of magnitudes larger than in the Raman case.
Maximal  entanglement ($NPT \ge 0.99$) is restricted only to the high intensities region ($\Omega_{p,c} \sim 6 \ \Gamma$) with  probabilities ${{\cal{P}}}$$\sim 10^{-4}$ provided that $\rho_{11}^0  \gg \rho_{22}^0$, \textit{i.e.}, most of the atomic background population~\cite{sm} is in the lowest state $\left\vert 1\right\rangle$.
Losses, on the other hand,  remain small  around the high intensity region due to the \textit{Autler-Townes} splitting and 
 in the lower intensities region due to \textit{Electromagnetically Induced Transparency} (\textit{EIT})~\cite{Scully:1997aa}.

In general, the atoms (optical) response depends in general on the atom's levels relative steady-state background atomic populations $~\varrho_{jj}$ with $\{j=1,2,3\}$ (see~\cite{no-excited}).
As there exists  widely used techniques for accurate state preparation of cold $^{85}$\textit{Rb} samples~\cite{Foot:2005aa} it may be worth to examine how entanglement may be engineered also through state preparation.

\begin{figure}[t!]
\includegraphics[width=8.5cm]{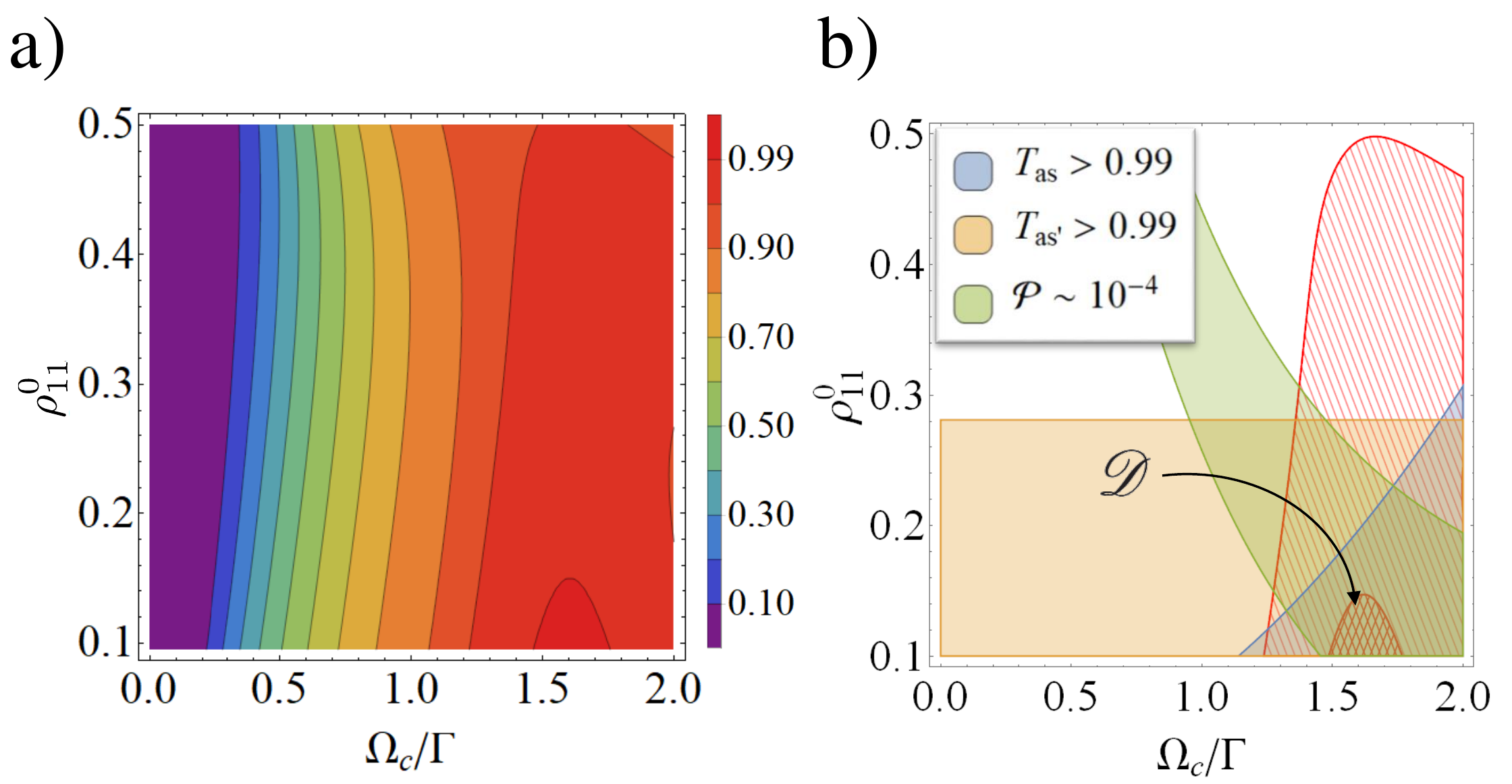} 
\caption{
{\it \textbf{Optimal  entanglement}} - \textit{\textbf{a}}) 
Same as in Fig.~\ref{fig:raman-eit} with $\Delta_p=-\Gamma$, $\Delta_c=\Gamma$,  $\Omega_p=1.5\ \Gamma$ and with varying background population values $\rho_{11}^0$ of the lowest level~\cite{no-excited}.
\textbf{\textit{b})}  
Regions where the $NPT\ge0.95$ (red line hatching) and $NPT\ge0.99$ (red cross hatching) as in Fig.~\ref{fig:eit2}a. 
Anti-Stokes transmission  ($T_{as}$) increases with $\Omega_c$ but strongly decreases with $\rho_{11}^0$ if $\Omega_c\lesssim \Gamma$.  
The other anti-Stokes transmission ($T_{as'}$) varies in the range 0.98$\div$1 for $0.5 > \rho_{11}^0 > 0.1$  and regardless of  $\Omega_c$.
The overlap region ($\mathscr{D}$) between the low absorption areas (light blue \& light orange) yield optimal conditions to generate color entanglement ($NPT\ge0.99$) with a probability ${\cal{P}} \sim 10^{-4}$.
\label{fig:eit2}
}
\end{figure}

Unlike the Raman-like regime, where the $NPT$ behaviour (Fig. 1) is unaffected by populations, 
changing the background population of the two lower levels $\ket{1}$ and $\ket{2}$ leads (see Fig.~\ref{fig:eit2}a) to optimal $NPT$ values  ($NPT\ge 0.99$) at  coupling strengths smaller than those of Fig.~\ref{fig:eit1}, with vanishing losses ($<1\%$). 
A plot with all figures of merit - $NPT, {\cal{P}}$ and absorption - is reported in Fig.~\ref{fig:eit2}b.
Optimal color entanglement is seen to take place within an overlap region for weak drivings $ \Omega_{p,c} \sim |\Delta_{p,c}| \sim \Gamma$ and a largely unbalanced background populations $\rho_{11}^0$ and $\rho_{22}^0$ of the two lower levels.

From Fig.~\ref{fig:eit2} it is clear how the manipulation of the lower populations yields to the optimization of  generation as it guarantees almost the same performances in terms of $NPT$ and $\cal{P}$ as those of Fig.~\ref{fig:eit1} yet  at much lower $\Omega_{c,p}$'s which provides an advantage toward implementation of the scheme with weak driving fields.

In conclusion, we anticipate that generation of heralded frequency-entangled single-photon states may occur through the Stokes state projection scheme of Fig.\ref{fig:exp-scheme}
whereby the detection of a single-Stokes photon in the frequency mode $\omega_s$ ($\omega_{s^\prime}$) does not reveal, even in principle, which frequency mode $\omega_{as}$ or $\omega_{as^\prime}$  is 
populated by a single-photon. 
This scenario provide new tools to develop quantum communication and information processing exploiting the frequency degree of freedom~\cite{Zavatta2014}. 
As is often the case for  non-classical effects of the electromagnetic field~\cite{Artoni:1994aa}, the proposal for entanglement generation through the indistinguishability between two different spontaneous four-wave mixing processes exhibits a certain versatility.
In principle, it could be adapted to atoms photonic crystal fibers interfaces~\cite{Venkataraman:2011ys, Bajcsy:2011yu},
to miniaturized (micrometer-sized) atomic vapour cells~\cite{Rost:2010aa,Baluktsian:2010aa} 
or to solid interfaces with crystals doped with rare-earth-metal ions~\cite{He:2006aa} or with  with N-V color centers~\cite{PhysRevB.77.113106, Doherty:2013uq}
where similar three-levels two-fold interaction configurations exist.

\textbf{Acknowledgments.}
We are indebted  to Prof. Jin Hui Wu for many insightful discussions. Support from the Italian Ministry of Foreign Affairs (MAECI) and the National Natural Science Foundation of China (NSFC) (Cooperative Program PGR00960), Ente Cassa di Risparmio di Firenze and the Italian Ministry of University and Research (MIUR) (QSecGroundSpace) are greatly acknowledged. 

\vspace{0.5cm}
$^*$ alessandro.zavatta@ino.it


\begin{thebibliography}{40}
\bibitem [{noA()}]{noAB}%
  \BibitemOpen
  \href@noop {} {}\bibinfo {note} {Conversely, the detection of a single-photon
  $\omega_{s^\prime}$ will prevent the processes $\{A,B\}$ from concurring to
  the generation of $\ket{\Psi}$.}\BibitemShut {Stop}%
\bibitem [{rel()}]{rel-phase}%
  \BibitemOpen
  \href@noop {} {}\bibinfo {note} {The relative phase of $f_A$ and $f_B$
  in~(\ref{final}) can be controlled by adjusting the pump and coupling
  relative phase.}\BibitemShut {Stop}%
\bibitem [{no-()}]{no-excited}%
  \BibitemOpen
  \href@noop {} {}\bibinfo {note} {The population difference
  $~\varrho_{jj}^0=\rho_{jj}^0-\rho_{33}^0 \simeq \rho_{jj}^0$ amounts to the
  steady-state background number density $\rho_{jj}^0=N_j^0/N$ that one has in
  the absence of pump and coupling and with $N \cong N_1^0+N_2^0$ being the
  total (conserved) number density.}\BibitemShut {Stop}%
\end{thebibliography}

%

\end{document}